\documentclass[preprint]{aastex}
\usepackage{natbib}
\usepackage{longtable}
\usepackage{lscape}
\slugcomment{draft version  \today}
\begin{document}
\title{The Small Numbers of Large Kuiper Belt Objects}
\author{Megan E. Schwamb\altaffilmark{1}\altaffilmark{2}\altaffilmark{3}\altaffilmark{4},Michael E. Brown\altaffilmark{4},and Wesley C. Fraser\altaffilmark{4,5}}
\altaffiltext{1}{Institute of Astronomy and Astrophysics, Academia Sinica, 11F of Astronomy-Mathematics Building, National Taiwan University. No.1, Sec. 4, Roosevelt Rd, Taipei 10617, Taiwan}
\altaffiltext{2}{Yale Center for Astronomy and Astrophysics, Yale University,P.O. Box 208121, New Haven, CT 06520}
\altaffiltext{3}{Department of Physics, Yale University, New Haven, CT 06511}
\altaffiltext{4}{Division of Geological and Planetary Sciences, California Institute of Technology, Pasadena, CA 91125}
\altaffiltext{5}{Herzberg Institute of Astrophysics, National Research Council, 5071 W. Saanich Road,Victoria, BC, V9E 2E7, Canada}

\email{mschwamb@asiaa.sinica.edu.tw}

\begin{abstract}

We explore the brightness distribution of the largest and brightest (\emph{m}(R)$<$22) Kuiper belt objects (KBOs). We construct a luminosity function of the dynamically excited  or hot Kuiper belt (orbits with inclinations $>$ 5$^{\circ}$) from the very brightest to \emph{m}(R)=23. We find for \emph{m}(R)$\lesssim$ 23, a single slope appears to describe the luminosity function. We estimate $\sim$12 KBOs brighter than \emph{m}(R)$\sim$19.5 are present in the Kuiper belt today. With 9 bodies already discovered this suggests that the inventory of bright KBOs is nearly complete. 
\end{abstract}

\section{Introduction}

The bodies residing in the Kuiper belt are the leftovers from the age of planet formation. The physical and orbital properties of these planetesimals serve as a record  of the Solar System's dynamical history and probe the conditions present in the primordial planetesimal disk. The size distribution of the Kuiper belt and its observational proxy - the luminosity function - are the end result of the accretional and collisional processes undergone during the creation and growth of these icy bodies. Exploring the size distribution of the Kuiper belt provides a unique test of and constraint on planetesimal formation theories  \citep{1999ApJ...526..465K, 2002PASP..114..265K, 2004AJ....128.1916K,2008ssbn.book..293K, 2010Icar..208..518C,2011ApJ...728...68S}. 

The size distribution of Kuiper belt objects (KBOs) has been studied most extensively for objects of moderate size \citep[e.g.][]{1998AJ....116.2042G, 2001AJ....121.1730L,2009AJ....137...72F, 2010arXiv1008.1058F,2010ApJ...722.1290F}, $22\le  \emph{m}(R) \le 25$ objects which can be found in relatively large numbers in modest surveys using medium-sized telescopes. For this brightness range, observations find that the luminosity function of the Kuiper belt is well represented by:
\begin{equation}
 \rm N( \le m)\simeq10^{ \alpha (m-m_0)}
 \end{equation}
where $N(\le$ m) is the cumulative number of objects per unit area brighter than or equal to magnitude
$m$, $\alpha$ is the logarithmic slope of the power law, and $m_0$ is the magnitude at which the sky density of objects with magnitude brighter than or equal to $m_o$ is 1 object per square degree measured on the ecliptic.
Values found for $\alpha$ range from 0.35 to 0.9  \citep[e.g.][]{1998AJ....115.2125J,1998AJ....116.2042G,2004AJ....128.1364B,2005AJ....129.1117E,2008Icar..198..452F,2008AJ....136...83F,2010arXiv1008.1058F}and are broadly consistent with
accretion theory  \citep{1999ApJ...526..465K,2002PASP..114..265K,2004AJ....128.1916K,2008ssbn.book..293K,2010Icar..208..518C, 2011ApJ...728...68S}.

While much attention has focused on the luminosity function of KBOs fainter than  \emph{m}(R) $\simeq25$, where a shallowing of the luminosity  function is a possible signature of collisional evolution of the Kuiper belt \citep{2004AJ....128.1364B,2008AJ....136...83F,2009AJ....137...72F,2010ApJ...722.1290F}, comparatively little attention has been paid to the luminosity function of the largest objects. Accretion models \citep{1999ApJ...526..465K,2002PASP..114..265K,2004AJ....128.1916K,2008ssbn.book..293K,2010Icar..208..518C,2011ApJ...728...68S} predict the slope of the KBO size distribution should continue smoothly to  the brightest object. Measurement of the luminosity function at the bright end (\emph{m}(R)$<$22) should thus provide strong constraints on these accretionary theories, but in fact, the luminosity function of the brightest KBOs is not as well known. This seemingly surprising situation is a result of the fact that in order to find the few large and bright (\emph{m}(R)$\lesssim$ 22) KBOs wide-field surveys \citep[such as][Rabinowitz et al. 2012]{2000AJ....120.2687S,2001AJ....121..562L,2003EM&P...92...99T,2005AJ....129.1117E,2007AJ....133.1247L,2008ssbn.book..335B,2010ApJ...720.1691S,2011AJ....142...98S} cover several thousands of square degrees over a wide range of conditions making precise photometric and detection efficiency calibration difficult. 

Previous attempts have been made at constructing the large KBO luminosity function. \cite{2008ssbn.book..335B} found the large KBOs (\emph{m}(R)$<$ 21) follow a single function with the same slope measured by \cite{2004AJ....128.1364B} at fainter magnitudes (smaller sizes), but made no attempt to calculate the detection efficiencies that would be required to absolutely calibrate the brightness distribution of the large KBOs and compare the absolute number of objects to that measured at smaller sizes. More recently \cite{2011AJ....142...98S} examined the cumulative number of KBOs as a function of absolute magnitude for all known KBOs and their survey discoveries but make no attempt to correct for detection losses and survey biases. Thus, the luminosity function of the brightest and largest KBOs (\emph{m}(R)$<$ 22) has not been properly joined with that observed at fainter magnitudes (smaller sizes) nor their absolute numbers compared.

\cite{2010ApJ...720.1691S} have provided the largest wide field survey to date with detections of these bright objects, moderately accurate photometric calibration, and an empirically  determined efficiency function. In this paper, we use  the \cite{2010ApJ...720.1691S} survey combined with available published surveys, to make the first attempt at constructing a complete luminosity function of the Kuiper belt from the brightest objects to \emph{m}(R)$\sim23$ and compare the brightness distribution obtained for the largest and brightest KBOs (\emph{m}(R)$<$22) to that measured for smaller fainter KBOs. 

\section{Data Sets}

No single survey to date has the sky coverage and depth to detect a sufficient number of objects for which the brightness distribution could be accurately measured over  $19\le \emph{m}(R) \le 25$ range. Accurately constructing the luminosity function across this magnitude range requires careful selection of comparison surveys to combine, objects within the surveys to include, and correction of each survey to a common system.

First we must select the objects to include in our luminosity function. Observations find that the luminosity function differs for dynamically cold classical KBOs (defined as \emph{i} $<$ 5$^\circ$ orbits) and dynamically excited or `hot' orbits (\emph{i} $>$ 5$^\circ$) \citep{2001AJ....121.1730L,2004AJ....128.1364B,2008AJ....136...83F,2010arXiv1008.1058F,2010ApJ...722.1290F}. The cold classicals are a set of exclusively red objects in low inclination, low eccentricity orbits with semimajor axes between about 42 and 48 AU, \citep{2004come.book..175M} that appear to be a physically distinct population with physical and dynamical characteristics (including color and binary fraction) distinct from the rest of the KBO population \citep{2002AJ....124.2279D,2004Icar..170..153P,2008Icar..194..758N,2008AJ....136.1837P}. The cold classical size range is also smaller than that of the dynamically excited KBOs  \citep{2001AJ....121.1730L}. The cold population lacks objects brighter than \emph{m}(R)=21.5, with nearly all the largest and brightest KBOs being members of the dynamically excited or hot KBO population. Therefore we restrict our analysis to the hot population only, ignoring the cold classical Kuiper belt.

For the analysis described in this Paper, we define our hot or excited KBO population, which we will refer to as the `hot population', as those objects with inclinations greater than 5 degrees and discovered at barycentric distances greater than 25 AU.   
We use the dynamical boundary at 5 degrees inclination found by \cite{2001AJ....121.2804B} to exclude the majority of  cold classical orbits. We note that there is observational evidence suggesting a break in the color distribution of the classical belt, separating red objects and more varied in color bodies, at a higher inclination of $\sim$12 degrees \citep{2008AJ....136.1837P}. The cause of  this discrepancy between the inclination distribution and the color distribution has yet to be resolved.The low-inclination peak due to the cold classicals in the inclination distribution is very well defined, and we therefore use this as the basis to remove cold classicals from the survey detections. While many surveys do not perform sufficient astrometric follow-up to precisely determine orbital parameters, even two-night observations are sufficient to determine the inclination and  of a minor planet to moderate accuracy. We have restricted our analysis to surveys where the majority of the detected KBOs have observed arcs of at least 24 hours in order to securely identify the hot KBOs with little contamination. We also exclude the much closer Centaur population from our luminosity function. While the Centaurs are derived from the Kuiper belt, their much closer distances would allow small objects to contaminate the  luminosity function of the brightest KBOs. Like inclination, heliocentric distance is also well-determined in short observations arcs.  We thus include in our sample only objects discovered at heliocentric distances greater than 25 AU where the  majority of the determined orbits will be beyond Neptune. 

Next, we must select appropriate surveys to combine in order to assemble our luminosity function. The number density of KBOs changes with ecliptic longitude. This variation is primarily due to the Plutinos, bodies residing in the 3:2 mean motion resonance with Neptune. A large concentration of Plutinos have orbits that come to perihelion at approximately 40-140 degrees ahead of and behind Neptune. Surveys observing at those ecliptic longitudes are biased towards the detection of these preferentially closer, thus brighter, objects. For deep pencil-beam surveys which search only a few square degrees over a narrow range of ecliptic longitudes this detection enhancement could be significant when binning into a cumulative magnitude distribution, especially for  \emph{m}(R)$<$ 22 where small numbers of non-resonant hot KBOs are expected. Then extrapolating from these surveys to full-sky would overestimate the number of KBOs as a function of magnitude. To properly account for these variations in sky density would require both simultaneously solving for the absolute magnitude and radial distribution for each of these surveys, which is beyond the scope of this paper. But for surveys searching several hundreds to thousands of square degrees, this effect is mitigated by the large swath of sky surveyed. They cover much more area where the majority of Plutinos are not coming to perihelia and biased towards detection. Plutinos will be a very small fraction of the surveys' overall detections, and thus correction is generally negligible for surveys of such sizes. When extrapolating to the full-sky, these wide-field surveys will do a much better job at reflecting the true numbers of KBOs. Thus we restrict our survey sample to those that cover $>$ 100 deg$^2$.

Table 1 summarizes the properties of each survey selected for the analysis presented here. While the \cite{2010ApJ...720.1691S} and \cite{2011arXiv1108.4836P}  surveys are the only wide-field surveys that include an estimate of an efficiency function, for comparison we also include the surveys of \cite{2003EM&P...92...99T} and \cite{2001AJ....121..562L} which have published detection lists and each found a significant number of objects brighter than \emph{m}(R)=21. For KBOs fainter than \emph{m}(R)=22, we use \cite{2011arXiv1108.4836P} which is the only survey that fits our selection criteria detecting moderately sized KBOs with a high recovery rate and well-characterized detection efficiency. In order to consistently calibrate the luminosity function, we restrict \cite{ 2011arXiv1108.4836P} to those observations within a few degrees of the ecliptic, excluding the two survey blocks observed at 10 and 20 degrees off the ecliptic. \cite{2011arXiv1108.4836P} determine the detection losses of each of their fields separately. We take their nominal survey efficiency to be the average detection efficiency of all the fields searched.
 
\section{The Hot KBO Luminosity Function}

In the following section we compute the hot KBO luminosity function and determine whether there is a match in sky density between 20$<$\emph{m}(R)$<$23 KBOs predicted by \cite{2011arXiv1108.4836P} to that measured at the the bright end, \emph{m}(R)$<$22,  from the \citep{2001AJ....121..562L,2003EM&P...92...99T,2010ApJ...720.1691S} surveys. We choose to assemble the luminosity function based upon apparent magnitude rather than absolute magnitude or estimate a size distribution (with an assumption for albedo). The detection efficiencies and limiting magnitudes for our sample surveys are all measured in terms of a flux limit. The absolute magnitude (or size) that a survey is sensitive to depends on the distance to the body. The hot population covers a much wider radial distance, ranging from 25 AU to approximately 100 AU, than the cold classicals where a mean distance of 42 AU is typically assumed to convert from the luminosity function to a size or absolute magnitude distribution. To correctly calibrate each survey in our sample  in terms of absolute magnitude would require knowing the full radial and orbital distribution of the hot population. Therefore we choose to avoid these complexities and simply use the apparent magnitudes which are a convolution of the hot population's radial, size, and albedo distributions.

Our survey sample (listed in Table 1) observes in a variety of different filters; we choose the R filter as our common magnitude reference system. For the \cite{2010ApJ...720.1691S} survey, each KBO was imaged four times, twice each night, and the apparent magnitude is taken to be the median of the four observations. We find an average $<$V$-$R$>$ color of 0.54  for multiopposition hot population KBOs (\emph{a} $>$  30 AU and \emph{i} $>$ 5$^\circ$) in the MBOSS Database\footnote{http://www.sc.eso.org/$\sim$ohainaut/MBOSS/} \citep{2002A&A...389..641H}. We use the magnitude transformation $<$g$^\prime$$-$ R$>$ =  0.8 used by \cite{2011arXiv1108.4836P} to transform their pre-survey detections R to g' . We apply these values as our constant offset to transform the reported survey apparent magnitudes to R band. The \cite{2003EM&P...92...99T} and \cite{2010ApJ...720.1691S} surveys both use the broadband RG610 filter, a broadband VR  filter. Using the magnitude transformations provided by \cite{2001ApJ...549L.241A}, we find a small average offset of $<$VR$-$R$>$= 0.02, and we choose to apply no offset to these surveys' reported magnitudes.  

For each survey, we compile the differential luminosity function, the number of KBOs as a function of apparent magnitude binned in 0.25 mag bins up to their limiting magnitudes. For \cite{2010ApJ...720.1691S}, we account for the survey losses by dividing by the reported detection efficiency in each magnitude bin. For \cite{2011arXiv1108.4836P}, we correct for detection losses by simply dividing by the  nominal survey detection efficiency. Additionally, \cite{2011arXiv1108.4836P} only report the objects that were successfully tracked in follow-up observations, and we account for their magnitude dependent recovery rate by dividing by the reported follow-up efficiency at each magnitude bin. As no efficiency is reported, we make no correction for \cite{2003EM&P...92...99T} and \cite{2001AJ....121..562L}. We then assemble the differential luminosity function into a cumulative distribution at 0.25 magnitude intervals. Error bars are then taken as the Poissonian 68$\%$ uncertainty (as prescribed by \citeauthor{1991ApJ...374..344K}  \citeyear{1991ApJ...374..344K}) for the value of the cumulative distribution in each magnitude bin. 

To directly compare the brightness distributions from each survey in our sample, we require a common reference sky coverage to account for observational biases caused by the on-sky density of KBOs varying with ecliptic latitude. We select the sky coverage of \cite{2010ApJ...720.1691S}, because the survey covers the most sky in our sample with published estimates of detection efficiency. \cite{2010ApJ...720.1691S} searched 11,786 deg$^2$ down to a mean limiting R magnitude of $\sim$21.3, within $\pm$30$^\circ$ of the ecliptic.  For further details about the survey and calibration, we refer the reader to \cite{2010ApJ...720.1691S}. 

For the remaining surveys in our sample, we calculate the number of objects that would have been found had they searched the same sky coverage as \cite{2010ApJ...720.1691S}. \cite{2010ApJ...720.1691S} targeted fields well off ecliptic, but  \cite{ 2001AJ....121..562L}, \cite{2003EM&P...92...99T}, and \cite{2011arXiv1108.4836P} observed fields at less than 10 degrees ecliptic latitude. If KBOs were uniformly distributed latitudinally, the total number of objects expected within the \cite{2010ApJ...720.1691S} survey region would simply be the number density found at the ecliptic multiplied by \cite{2010ApJ...720.1691S}'s areal coverage (11,786 deg$^2$). However, the number of KBOs varies as a function of distance from the ecliptic, and we must account for that.

\begin{figure}
\epsscale{1}
\plotone{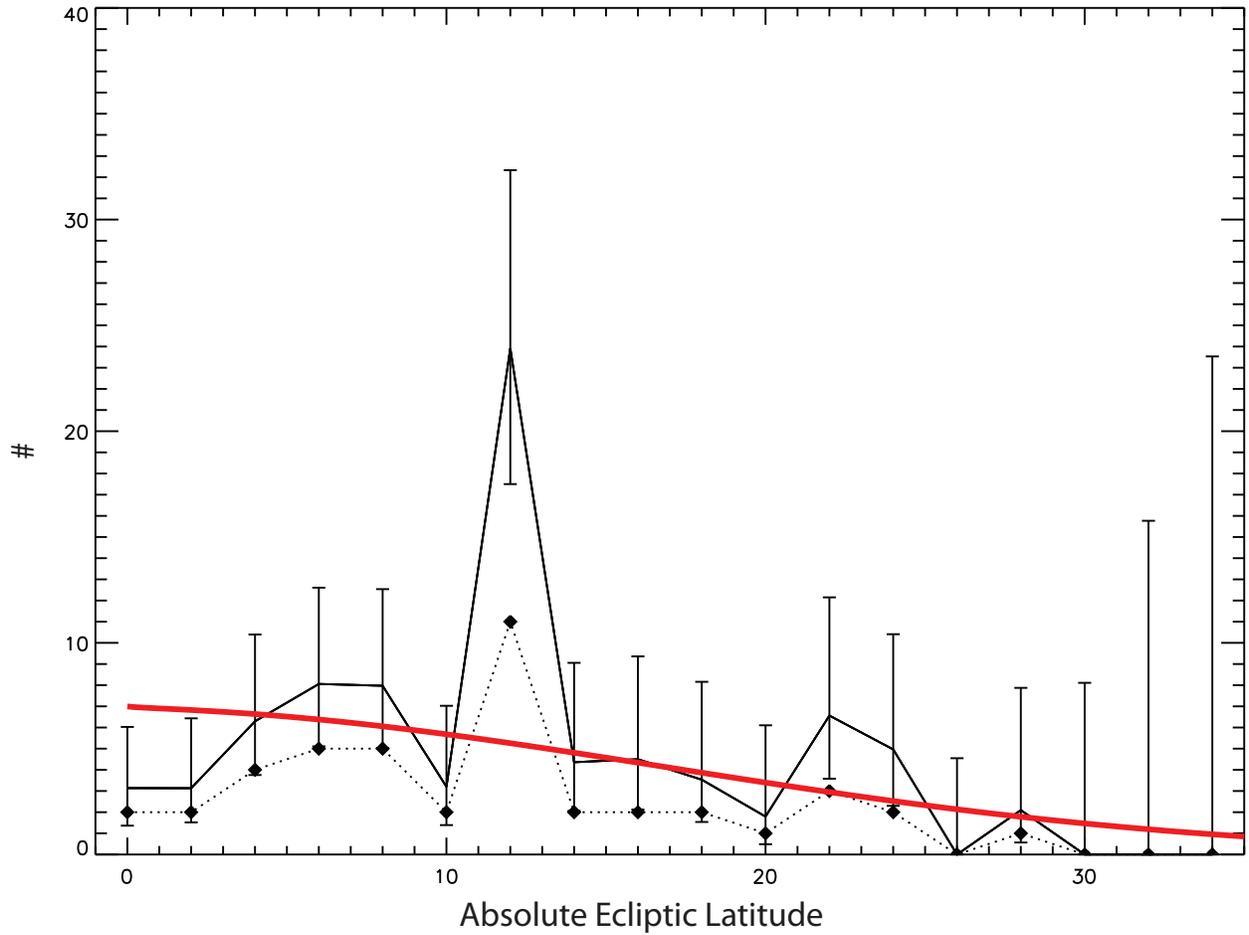}
\caption{ The folded latitudinal distribution of objects found in \cite{2010ApJ...720.1691S}. The lower dashed line with diamonds shows the number of actual KBO detections in two-degree bins. The solid line shows the expected number of KBOs brighter than 21.3 corrected for sky coverage with one-$\sigma$ Poisson error bars computed for the unfolded distribution added in quadrature. The best-fit latitude distribution is plotted in red (online version)}
\label{fig:lat}
\end{figure}

We compute the correction from a flat distribution using the latitude distribution derived, by \cite{2001AJ....121.2804B} scaled to fit the observed \cite{2010ApJ...720.1691S} folded latitude distribution. Making an approximation for circular orbits, L($\beta$), the number of KBOs as a function of absolute ecliptic latitude ($\beta$) is:
 \begin{equation}
{\rm L}(\beta) \propto \int_0^{\pi}  \frac{\,{\rm cos}(\beta) e^{ \left({\frac{-i^2}{2\sigma^2}}\right)}\sin({i)}}{\sqrt{{\rm sin^2}(i)+{\rm sin^2}(\beta)}} \,{\rm d}i
\end{equation} 
where $i$ is inclination,$\sigma$ is 15$^\circ$for the hot KBOs. The \cite{2010ApJ...720.1691S} observed folded latitude distribution binned in 2 degree bins and ${\rm L}(\beta)$ scaled to the observed distribution are plotted in Figure \ref{fig:lat}. \cite{2010ApJ...720.1691S} observes a spike in detections at  $\sim\pm$10 degrees that is not present in ${\rm L}(\beta)$. These peaks were hypothesized by \cite{2008ssbn.book..335B} and \cite{2010ApJ...720.1691S} to be enhanced detections of the subset of Plutinos  locked in the Kozai resonance. With only 3 of the 33 \cite{2010ApJ...720.1691S} sample KBOs identified as potential Kozai Plutinos and modeling  of the Kozai Plutino population by \cite{2013AJ....146....6L} not reproducing a spike at those ecliptic latitudes, the excess in the 11-13 degree bin may be due to small-number statistics.  We thus chose to use the  \citet*{2001AJ....121.2804B} latitude distribution shown in Figure \ref{fig:lat} without an additional component at 11-13 degrees.

We find that the total number of objects in the area of the \cite{2010ApJ...720.1691S} survey should scale as the number density found at the ecliptic multiplied by the areal coverage of the \cite{2010ApJ...720.1691S} survey multiplied by a scaling factor of 0.69. While the precise value of this scaling is uncertain, the scaling itself is modest, so the precise value has only a small effect on the final results. The \cite{ 2001AJ....121..562L}, \cite{2003EM&P...92...99T}, and  \cite{2011arXiv1108.4836P}  density measurements are not strictly equatorial, but the latitude distribution below 10 degrees is essentially flat, so to transform these distributions to the \cite{2010ApJ...720.1691S} sky coverage, we scale for the difference in sky coverage between \cite{2010ApJ...720.1691S} and the respective surveys and apply the standard 0.69 scaling factor to the distribution and uncertainties in each magnitude bin.  We note the errors induced by the assumed hot population latitude distribution will affect all the corrected brightness distributions equally and therefore  not change the relative difference between these surveys.

 \section{Luminosity Function of the Bright KBOs}

\begin{figure}
\epsscale{1}
\plotone{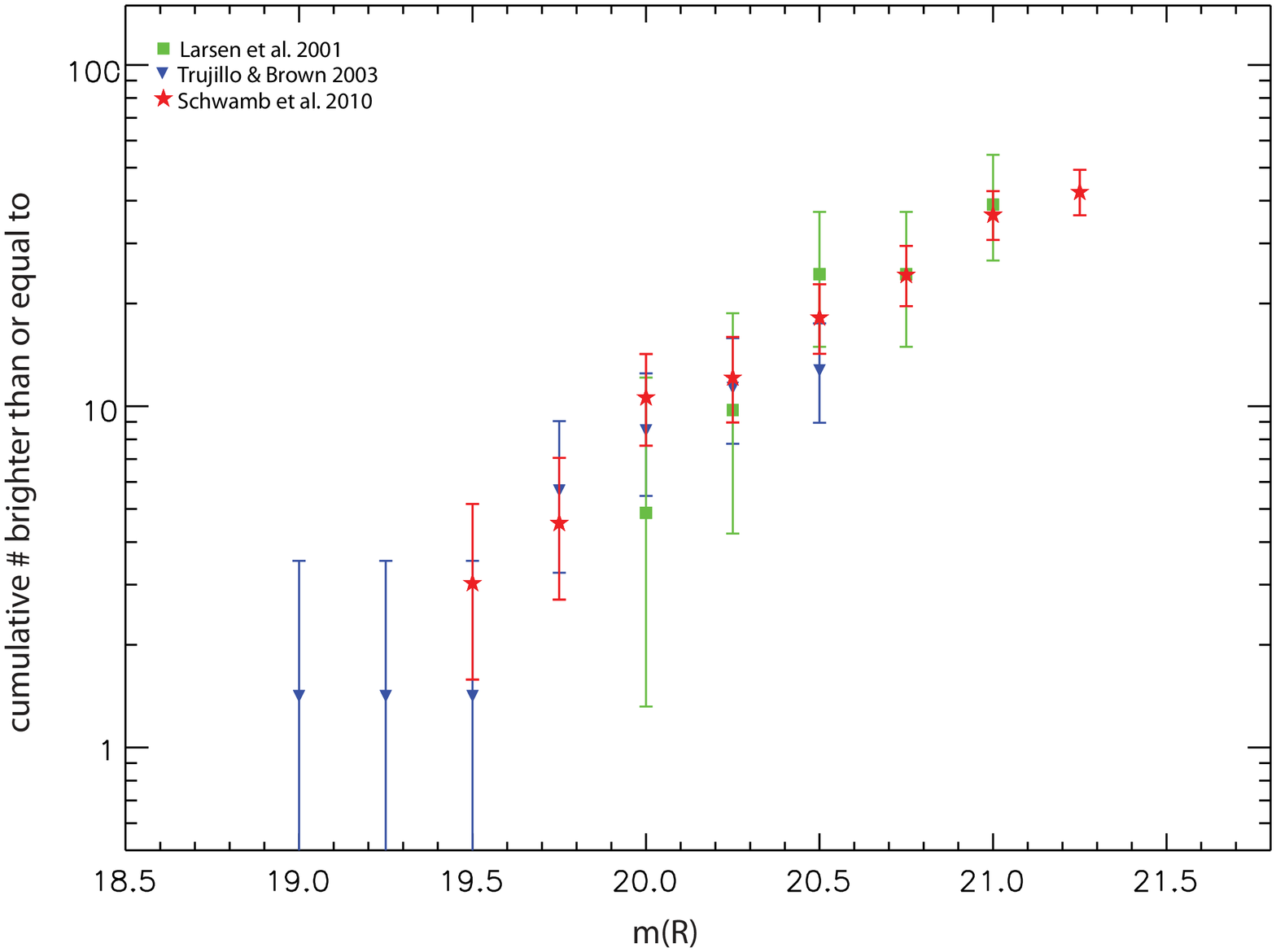}
\caption{ Cumulative distribution of hot population KBOs with R magnitude brighter than or equal expected to be present within the  \cite{2010ApJ...720.1691S}  survey region for the three sample wide-field surveys with limiting magnitudes less than \emph{m}(R)=22  \citep{2003EM&P...92...99T,2001AJ....121..562L,2010ApJ...720.1691S} calculated at 0.25 mag intervals. The plotted error bars represent the Poissonian 68$\%$ uncertainty for each magnitude bin. Note that the error bars for each bin correlated with those from  brighter bins.}
\label{fig:wide}
\end{figure}

We first examine the brightness distribution of the large KBOs ( \emph{m}(R)$<$ 22) obtained from \cite{2001AJ....121..562L},\cite{2003EM&P...92...99T}, and \cite{2010ApJ...720.1691S} . Figure \ref{fig:wide}  plots the number of hot population KBOs brighter than or equal to a given apparent R magnitude present in the \cite{2010ApJ...720.1691S} survey region binned in 0.25 mag bins for all three shallow wide-field surveys. All three surveys are photometrically calibrated to the USNO catalogs and the uncertainty in their measured magnitudes is approximately $\pm$ 0.3 mag \citep{1998usno.book.....M}, which could at most shift the cumulative luminosity function by a magnitude bin in either direction. Both \cite{2001AJ....121..562L} and \cite{2003EM&P...92...99T}  distributions are within a factor of 1.6 of the \cite{2010ApJ...720.1691S} distribution. The surveys are in relatively good agreement despite the \cite{2001AJ....121..562L} and \cite{2003EM&P...92...99T} surveys having no estimate of detection losses and the uncertainty in the estimated \cite{2010ApJ...720.1691S} survey efficiency. The corrections for these two surveys would likely be not more than a factor of two, and any correction would further improve the match between the luminosity functions. While we do not include the \cite{2008ssbn.book..335B} survey (of which the \citeauthor{2003EM&P...92...99T}  \citeyear{2003EM&P...92...99T} detections are a subset) because of a lack of a published detection list, we nonetheless note that the shape of the cumulative number of objects agrees well with this survey. 

 We can estimate the total number of large bodies (\emph{R}$\leq$19.5) in the Kuiper belt using the known objects reported to the Minor Planet Center\footnote{ http://www.cfa.harvard.edu/iau/Ephemerides/Distant/index.html} (MPC). The bulk of the \cite{2010ApJ...720.1691S}  sky coverage is within 30 degrees of the ecliptic where the majority of the hot population objects are found (see Figure 1), therefore we can estimate the number of bright KBOs visible in the Kuiper belt. The brightest body in the \cite{2010ApJ...720.1691S} survey is Quaoar with an \emph{m}(R) $\simeq$19 reported in the MPC. Four hot KBOs brighter than Quaoar (including Pluto Eris, Haumea, and Makemake) are known in the MPC. Scaling the \cite{2010ApJ...720.1691S} distribution, we find that approximately 12 hot KBOs brighter than or equal to 19.5 R mag are present within the Kuiper belt today. Nine  \emph{m}(R)$\leq$19.5 hot KBOs have previously been found and reported to the MPC. This suggests that the majority of the brightest KBOs have already been discovered, with perhaps 1 or 2 remaining to be found in the galactic plane or southern hemisphere. New surveys searching regions of the southern hemisphere not surveyed previously to \emph{m}(R)=21 have yet to find a new \emph{m}(R)$<$19.5 KBO  \citep[Rabinowitz et al 2012,][]{2011AJ....142...98S}. 

\section{Combined Luminosity Function}

\begin{figure}
\epsscale{1}
\plotone{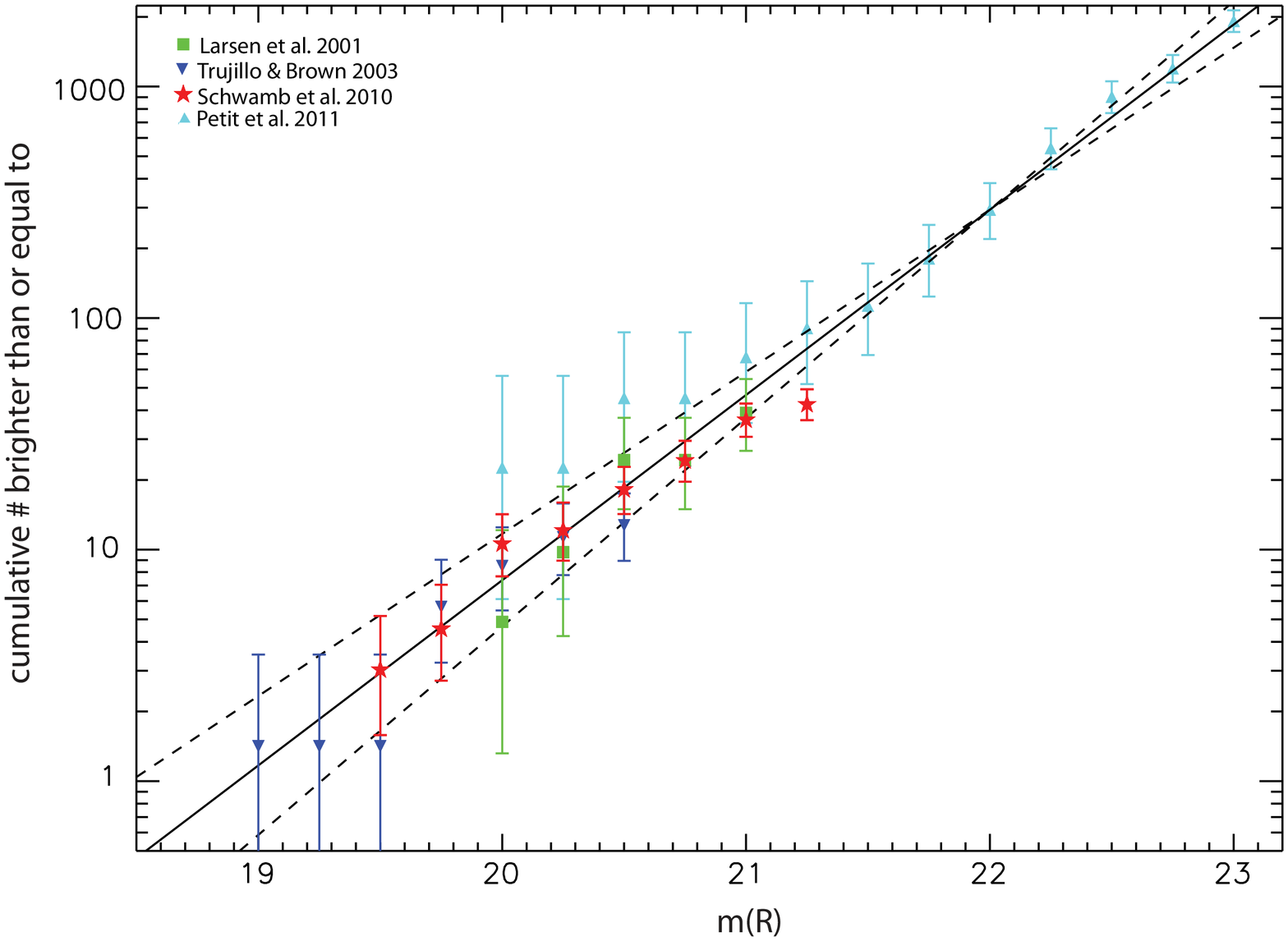}
\caption{ Combined cumulative luminosity of hot population KBOs with R magnitude ( \emph{m}(R)) brighter than or equal expected to be present within the \cite{2010ApJ...720.1691S}  survey region for \citep{2003EM&P...92...99T,2001AJ....121..562L,2010ApJ...720.1691S} and  \cite{2011arXiv1108.4836P} calculated at 0.25 mag intervals. The plotted error bars represent the Poissonian 68$\%$ uncertainty for each magnitude bin.  Note that the error bars for each bin are correlated with those from  brighter bins. The solid line plots the best-fit slope of  $\alpha$=0.81 (the solid line) from \cite{2011arXiv1108.4836P}  scaled to the value at  \emph{m}(R)=22  The dashed lines represent  $\alpha$=0.7 and 0.9 respectively, the approximate 1-$\sigma$ errors uncertainties from \cite{2011arXiv1108.4836P}. }
\label{fig:all}
\end{figure}

Figure \ref{fig:all} shows the full cumulative luminosity function within the \cite{2010ApJ...720.1691S} sky coverage from 19 $\le$ \emph{m}(R)$\le$ 23 including the results from \cite{2011arXiv1108.4836P}.  A single luminosity function can be found which, within the uncertainties, fits the entire combined survey set.  \cite{2011arXiv1108.4836P} measure the luminosity function for a dynamical subgroup of the hot population, finding a slope of $\alpha$=0.81 for the hot classical belt (those KBOs residing on fairly circular orbits within $\sim$42 to 48 AU with inclinations greater than 5 degrees).  In Figure \ref{fig:all} we plot this best-fit slope (solid line) scaled to the value \cite{2011arXiv1108.4836P} distribution \emph{m}(R)=22. A different value could have been chosen, but selected a value sufficiently far from the limiting magnitude and far enough from the bright end where small number of detections produce large uncertainties.  We also plot  $\alpha$=0.9 and  $\alpha$=0.7 for reference (dashed lines), the approximate 1-$\sigma$ errors uncertainties from \cite{2011arXiv1108.4836P}.  We find a slope of $\alpha$=0.81 well describes the luminosity function of the hot KBOs for \emph{m}(R)$<$ 23, the same slope found by \cite{2011AJ....142...98S} for H $<7$ KBOs and slightly steeper than the hot population slope measured by \cite{2005AJ....129.1117E} for 20 $\leq$ m(R)$\leq$22.5. The number of bright KBOs expected from \emph{m}(R)$<$ 23 estimates are consistent with the shallow wide-field survey \citep{2001AJ....121..562L,2003EM&P...92...99T,2010ApJ...720.1691S} detections.

Although not in our sample,  \cite{2010arXiv1008.1058F}  is currently the only other survey with a sufficient sample of objects detected over a range of magnitudes to measure a luminosity function independently for  22 $\leq$ \emph{m}(R)$<$ 25  without the inherit biases or issues from combining multiple survey detections. Covering only 8 deg$^2$, we are  unable to effectively calibrate \cite{2010arXiv1008.1058F}  to the \cite{2010ApJ...720.1691S} sky coverage and include the survey in the analysis presented here. \cite{2010arXiv1008.1058F} find the hot KBO luminosity function is well fit by a single slope, measuring a relatively flat slope of $\alpha$=0.35$\pm$0.21 for the hot population, far shallower than the \cite{2011arXiv1108.4836P} slope of $\alpha$=0.81$^{+0.3}_{-0.2}$ at the 95$\%$ confidence level  that is consistent for R $<$ 23  hot KBOs. \cite{2010arXiv1008.1058F} specifically observed at  longitudes where the Plutinos preferentially come to perihelia away from Neptune. In \cite{2010arXiv1008.1058F}'s sample, objects found at 30$<d<$38 AU will be a mixture of primarily Plutinos and non-resonant hot KBOs. On the other hand, non-Plutino orbits will dominate detections at 38$<d<$55 AU. \cite{2010arXiv1008.1058F} find the same shallow slope for both objects within the closer  30$<d<$38 AU and more distant 38$<d<$55 AU samples suggesting this is not an effect caused by detecting far more closer, therefore smaller Plutinos than \cite{2011arXiv1108.4836P}.  If the nominal slopes measured by \cite{2010arXiv1008.1058F} and \cite{2011arXiv1108.4836P} are correct, at magnitudes fainter than \emph{m}(R)=23,  the luminosity function of the hot KBOs would have to transition to a shallower slope in order to accommodate the $\alpha$=0.35 measured by  \cite{2010arXiv1008.1058F}.  Although  \cite{2010ApJ...722.1290F} suffers from the effects of combining multiple survey detections, their results  support a change of slope for the hot population luminosity function at magnitudes  fainter  than \emph{m}(R)=23.  \cite{2010ApJ...722.1290F} combined Hubble Space Telescope discoveries of hot KBOs fainter than 25th magnitude  with shallower surveys, finding  the hot population luminosity  function transitions at  R magnitude of 24.1$\pm$0.7 to a slope of  $\alpha$=0.30$\pm$0.07.  Further observations are required to confirm the exact shape of the hot population luminosity function at  magnitudes fainter than 23rd and confirm this break in the luminosity function.

\section{Conclusions}

Combining observations from available wide-field and deep surveys we make the first attempt at constructing a complete luminosity function of the dynamically excited Kuiper belt from the brightest objects to $R$$\sim$23. Comparing the brightness distribution obtained for the largest and brightest KBOs from  the  \cite{2001AJ....121..562L},\cite{2003EM&P...92...99T}, and \cite{2010ApJ...720.1691S}  surveys to that measured for smaller fainter KBOs by  \cite{2011arXiv1108.4836P}, we find for \emph{m}(R)$<$23, a single slope luminosity function  describes the hot population luminosity function. Both the number and slope of the distributions match. We estimate $\sim$12 dynamically hot KBOs brighter than \emph{m}(R)$\simeq$19.5  are present in the Kuiper belt today implying the inventory of bright KBOs is almost complete.

For  \emph{m}(R)$>$23,  a single slope brightness distribution may not be sufficient to describe the luminosity function of the hot population. \cite{2011arXiv1108.4836P}  is most sensitive to measuring the luminosity function in the 22-23.5 R magnitude range, and  \cite{2010arXiv1008.1058F} probes magnitudes from  23-25.0. If the nominal slopes for both surveys are correct, it appears that the steeper slope of $\alpha$=0.81$^{+0.3}_{-0.2}$  slope measured  \cite{2011arXiv1108.4836P}  transitions to a shallower slope at R magnitudes fainter than 23  in order to accommodate the $\alpha$=0.35 $\pm$0.21 slope measured by \cite{2010arXiv1008.1058F}. With the current set of observations, with our analysis, we cannot examine the brightness distribution fainter than 23rd magnitude. A complete picture of the hot population luminosity function is in need and requires further observations with sufficient numbers of objects from $22 \le  \emph{m}(R) \le 25$ to confirm this changeover and probe the exact nature of the hot population luminosity function for KBOs fainter than  \emph{m}(R)=23.

\noindent \emph{Acknowledgements} This research was supported by NASA Origins of Solar Systems Program grant NNG05GI02G. MES was supported in part by a NASA Earth and Space Science Fellowship and by an NSF Astronomy and Astrophysics Postdoctoral Fellowship under award AST-1003258. MES is currently supported by  an Academia Sinica Postdoctoral Fellowship.  The authors especially thank the referee for the careful and constructive feedback that  significantly improved the contents of this paper. We thank Chris Lintott, Brooke Simmons, Luke Dones, Andy Szymkowiak, and Darin Ragozzine for thoughtful and detailed manuscript comments and suggestions.

\bibliographystyle{apj}

\begin{thebibliography}{36}
\expandafter\ifx\csname natexlab\endcsname\relax\def\natexlab#1{#1}\fi

\bibitem[{{Allen} {et~al.}(2001){Allen}, {Bernstein}, \&
  {Malhotra}}]{2001ApJ...549L.241A}
{Allen}, R.~L., {Bernstein}, G.~M., \& {Malhotra}, R. 2001, \apjl, 549, L241

\bibitem[{{Bernstein} {et~al.}(2004){Bernstein}, {Trilling}, {Allen}, {Brown},
  {Holman}, \& {Malhotra}}]{2004AJ....128.1364B}
{Bernstein}, G.~M., {Trilling}, D.~E., {Allen}, R.~L., {et~al.} 2004, \aj, 128,
  1364

\bibitem[{{Brown}(2001)}]{2001AJ....121.2804B}
{Brown}, M.~E. 2001, \aj, 121, 2804

\bibitem[{{Brown}(2008)}]{2008ssbn.book..335B}
---. 2008, The Solar System Beyond Neptune, ed. M.~A. Barucci et al. (Tucson,
  AZ: Univ. of Arizona Press), 335

\bibitem[{{Cuzzi} {et~al.}(2010){Cuzzi}, {Hogan}, \&
  {Bottke}}]{2010Icar..208..518C}
{Cuzzi}, J.~N., {Hogan}, R.~C., \& {Bottke}, W.~F. 2010, \icarus, 208, 518

\bibitem[{{Doressoundiram} {et~al.}(2002){Doressoundiram}, {Peixinho}, {de
  Bergh}, {Fornasier}, {Th{\'e}bault}, {Barucci}, \&
  {Veillet}}]{2002AJ....124.2279D}
{Doressoundiram}, A., {Peixinho}, N., {de Bergh}, C., {et~al.} 2002, \aj, 124,
  2279

\bibitem[{{Elliot} {et~al.}(2005){Elliot}, {Kern}, {Clancy}, {Gulbis},
  {Millis}, {Buie}, {Wasserman}, {Chiang}, {Jordan}, {Trilling}, \&
  {Meech}}]{2005AJ....129.1117E}
{Elliot}, J.~L., {Kern}, S.~D., {Clancy}, K.~B., {et~al.} 2005, \aj, 129, 1117

\bibitem[{{Fraser} {et~al.}(2010){Fraser}, {Brown}, \&
  {Schwamb}}]{2010arXiv1008.1058F}
{Fraser}, W.~C., {Brown}, M.~E., \& {Schwamb}, M.~E. 2010, \icarus, 210, 944

\bibitem[{{Fraser} \& {Kavelaars}(2008)}]{2008Icar..198..452F}
{Fraser}, W.~C., \& {Kavelaars}, J.~J. 2008, \icarus, 198, 452

\bibitem[{{Fraser} \& {Kavelaars}(2009)}]{2009AJ....137...72F}
---. 2009, \aj, 137, 72

\bibitem[{{Fuentes} \& {Holman}(2008)}]{2008AJ....136...83F}
{Fuentes}, C.~I., \& {Holman}, M.~J. 2008, \aj, 136, 83

\bibitem[{{Fuentes} {et~al.}(2010){Fuentes}, {Holman}, {Trilling}, \&
  {Protopapas}}]{2010ApJ...722.1290F}
{Fuentes}, C.~I., {Holman}, M.~J., {Trilling}, D.~E., \& {Protopapas}, P. 2010,
  \apj, 722, 1290

\bibitem[{{Gladman} {et~al.}(1998){Gladman}, {Kavelaars}, {Nicholson},
  {Loredo}, \& {Burns}}]{1998AJ....116.2042G}
{Gladman}, B., {Kavelaars}, J.~J., {Nicholson}, P.~D., {Loredo}, T.~J., \&
  {Burns}, J.~A. 1998, \aj, 116, 2042

\bibitem[{{Hainaut} \& {Delsanti}(2002)}]{2002A&A...389..641H}
{Hainaut}, O.~R., \& {Delsanti}, A.~C. 2002, \aap, 389, 641

\bibitem[{{Jewitt} {et~al.}(1998){Jewitt}, {Luu}, \&
  {Trujillo}}]{1998AJ....115.2125J}
{Jewitt}, D., {Luu}, J., \& {Trujillo}, C. 1998, \aj, 115, 2125

\bibitem[{{Kenyon}(2002)}]{2002PASP..114..265K}
{Kenyon}, S.~J. 2002, \pasp, 114, 265

\bibitem[{{Kenyon} \& {Bromley}(2004)}]{2004AJ....128.1916K}
{Kenyon}, S.~J., \& {Bromley}, B.~C. 2004, \aj, 128, 1916

\bibitem[{{Kenyon} {et~al.}(2008){Kenyon}, {Bromley}, {O'Brien}, \&
  {Davis}}]{2008ssbn.book..293K}
{Kenyon}, S.~J., {Bromley}, B.~C., {O'Brien}, D.~P., \& {Davis}, D.~R. 2008,
  {Formation and Collisional Evolution of Kuiper Belt Objects}, 293--313

\bibitem[{{Kenyon} \& {Luu}(1999)}]{1999ApJ...526..465K}
{Kenyon}, S.~J., \& {Luu}, J.~X. 1999, \apj, 526, 465

\bibitem[{{Kraft} {et~al.}(1991){Kraft}, {Burrows}, \&
  {Nousek}}]{1991ApJ...374..344K}
{Kraft}, R.~P., {Burrows}, D.~N., \& {Nousek}, J.~A. 1991, \apj, 374, 344

\bibitem[{{Larsen} {et~al.}(2001){Larsen}, {Gleason}, {Danzl}, {Descour},
  {McMillan}, {Gehrels}, {Jedicke}, {Montani}, \&
  {Scotti}}]{2001AJ....121..562L}
{Larsen}, J.~A., {Gleason}, A.~E., {Danzl}, N.~M., {et~al.} 2001, \aj, 121, 562

\bibitem[{{Larsen} {et~al.}(2007){Larsen}, {Roe}, {Albert}, {Descour},
  {McMillan}, {Gleason}, {Jedicke}, {Block}, {Bressi}, {Cochran}, {Gehrels},
  {Montani}, {Perry}, {Read}, {Scotti}, \& {Tubbiolo}}]{2007AJ....133.1247L}
{Larsen}, J.~A., {Roe}, E.~S., {Albert}, C.~E., {et~al.} 2007, \aj, 133, 1247

\bibitem[{{Lawler} \& {Gladman}(2013)}]{2013AJ....146....6L}
{Lawler}, S.~M., \& {Gladman}, B. 2013, \aj, 146, 6

\bibitem[{{Levison} \& {Stern}(2001)}]{2001AJ....121.1730L}
{Levison}, H.~F., \& {Stern}, S.~A. 2001, \aj, 121, 1730

\bibitem[{{Monet}(1998)}]{1998usno.book.....M}
{Monet}, D. 1998, {USNO-A2.0} (USNO-A2.0, by Monet, David.~ [Flagstaff, AZ] :
  U.S.~Naval Observatory, c1998.~.~United States Naval Observatory.)

\bibitem[{{Morbidelli} \& {Brown}(2004)}]{2004come.book..175M}
{Morbidelli}, A., \& {Brown}, M.~E. 2004, {The kuiper belt and the primordial
  evolution of the solar system}, ed. {Festou, M.~C., Keller, H.~U., \& Weaver,
  H.~A.}, 175--191

\bibitem[{{Noll} {et~al.}(2008){Noll}, {Grundy}, {Stephens}, {Levison}, \&
  {Kern}}]{2008Icar..194..758N}
{Noll}, K.~S., {Grundy}, W.~M., {Stephens}, D.~C., {Levison}, H.~F., \& {Kern},
  S.~D. 2008, \icarus, 194, 758

\bibitem[{{Peixinho} {et~al.}(2004){Peixinho}, {Boehnhardt}, {Belskaya},
  {Doressoundiram}, {Barucci}, \& {Delsanti}}]{2004Icar..170..153P}
{Peixinho}, N., {Boehnhardt}, H., {Belskaya}, I., {et~al.} 2004, \icarus, 170,
  153

\bibitem[{{Peixinho} {et~al.}(2008){Peixinho}, {Lacerda}, \&
  {Jewitt}}]{2008AJ....136.1837P}
{Peixinho}, N., {Lacerda}, P., \& {Jewitt}, D. 2008, \aj, 136, 1837

\bibitem[{{Petit} {et~al.}(2011){Petit}, {Kavelaars}, {Gladman}, {Jones},
  {Parker}, {Van Laerhoven}, {Nicholson}, {Mars}, {Rousselot}, {Mousis},
  {Marsden}, {Bieryla}, {Taylor}, {Ashby}, {Benavidez}, {Campo Bagatin}, \&
  {Bernabeu}}]{2011arXiv1108.4836P}
{Petit}, J.-M., {Kavelaars}, J.~J., {Gladman}, B.~J., {et~al.} 2011, \aj, 142,
  131

\bibitem[{{Schlichting} \& {Sari}(2011)}]{2011ApJ...728...68S}
{Schlichting}, H.~E., \& {Sari}, R. 2011, \apj, 728, 68

\bibitem[{{Schwamb} {et~al.}(2010){Schwamb}, {Brown}, {Rabinowitz}, \&
  {Ragozzine}}]{2010ApJ...720.1691S}
{Schwamb}, M.~E., {Brown}, M.~E., {Rabinowitz}, D.~L., \& {Ragozzine}, D. 2010,
  \apj, 720, 1691

\bibitem[{{Sheppard} {et~al.}(2000){Sheppard}, {Jewitt}, {Trujillo}, {Brown},
  \& {Ashley}}]{2000AJ....120.2687S}
{Sheppard}, S.~S., {Jewitt}, D.~C., {Trujillo}, C.~A., {Brown}, M.~J.~I., \&
  {Ashley}, M.~C.~B. 2000, \aj, 120, 2687

\bibitem[{{Sheppard} {et~al.}(2011){Sheppard}, {Udalski}, {Trujillo}, {Kubiak},
  {Pietrzynski}, {Poleski}, {Soszynski}, {Szyma{\'n}ski}, \&
  {Ulaczyk}}]{2011AJ....142...98S}
{Sheppard}, S.~S., {Udalski}, A., {Trujillo}, C., {et~al.} 2011, \aj, 142, 98

\bibitem[{{Smith} {et~al.}(2002){Smith}, {Tucker}, {Kent}, {Richmond},
  {Fukugita}, {Ichikawa}, {Ichikawa}, {Jorgensen}, {Uomoto}, {Gunn}, {Hamabe},
  {Watanabe}, {Tolea}, {Henden}, {Annis}, {Pier}, {McKay}, {Brinkmann}, {Chen},
  {Holtzman}, {Shimasaku}, \& {York}}]{2002AJ....123.2121S}
{Smith}, J.~A., {Tucker}, D.~L., {Kent}, S., {et~al.} 2002, \aj, 123, 2121

\bibitem[{{Trujillo} \& {Brown}(2003)}]{2003EM&P...92...99T}
{Trujillo}, C.~A., \& {Brown}, M.~E. 2003, Earth Moon and Planets, 92, 99

\end{thebibliography}

\footnotesize
\begin{table}
\centering
\footnotesize
\begin{tabular}{ l c c l l l l}
\hline
\hline
  &  total $ \#$  hot  $^a$ & $\#$ hot KBOs  $^a$ & ecliptic & sky coverage & limiting & detection \\
Survey & KBOs found & mag $\le$ &  latitude (deg)  & (deg$^2$) &  magnitude & efficiency $\%$ \\ 
& & limiting magnitude & & & & \\

Larsen et al. (2001) & 9 & 8 & $<$10 & 1483.8 &  21.3 (21.8 V) & NA \\
Trujillo $\&$ Brown (2003) & 26 & 12 & $\pm$ 10 &  5108 & 20.7 (20.7  RG610) & NA \\
\cite{2010ApJ...720.1691S} & 44 & 30 &  $\pm$  40 &  11786 & 21.3 (21.3 RG610) & 66.0 \\
Petit et al. (2011) & 86 & 77 & $<$ 5 & 299 & 23.2 R (24 g$^\prime$) & 86.0  \\
 \\
 \hline
\hline

\hline
\hline
\end{tabular}
 \caption{ Kuiper belt surveys used in our analysis. $^a$  \emph{r} $>$25 AU and \emph{i} $>$ 5$^\circ$. NA= Not Available }
\label{tab:alltogether}
\end{table}

\normalsize

\end{document}